 \documentclass[12pt,preprint]{aastex}

\slugcomment{Accepted for publication in ApJL}

\shorttitle{Millimeter observations of obscured Spitzer sources}
\shortauthors{Lutz et al.}

\begin{document}

\title{Millimeter observations of obscured Spitzer 24 micron sources}


\author{D. Lutz\altaffilmark{1}, 
L. Yan \altaffilmark{2}, 
L. Armus \altaffilmark{2},
G. Helou \altaffilmark{2},
L.J. Tacconi \altaffilmark{1},
R. Genzel \altaffilmark{1},
A.J. Baker \altaffilmark{3,4}} 
\altaffiltext{1}{Max-Planck-Institut f\"ur extraterrestrische Physik, 
Postfach 1312, 85741 Garching, Germany \email{lutz@mpe.mpg.de, 
linda@mpe.mpg.de, genzel@mpe.mpg.de} }
\altaffiltext{2}{Spitzer Science Center, California Institute of 
Technology, 1200 East California Boulevard, MC 220-6, Pasadena, 
CA 91125 \email{lyan@ipac.caltech.edu, lee@ipac.caltech.edu, 
gxh@ipac.caltech.edu}}
\altaffiltext{3}{Jansky Fellow, National Radio Astronomy Observatory}
\altaffiltext{4}{Department of Astronomy, University of Maryland, 
College Park, MD 20742-2421 \email{ajb@astro.umd.edu} }

\begin{abstract}
We present MAMBO 1.2mm observations of 40 extragalactic sources from the 
Spitzer First Look 
Survey that are bright in the mid-infrared (S$_{24\mu m}>1$mJy) but 
optically obscured ($log_{10}(\nu F_\nu (24\mu m)/\nu F_\nu (0.7\mu m))>1$).
We use these observations to search for cold dust emission, probing
the similarity of their spectral energy distributions to star forming infrared 
galaxies or obscured AGN.
The sample as a whole is well detected at mean S$_{1.2mm}=0.74\pm0.09$mJy and
S$_{1.2mm}$/S$_{24\mu m}=0.15\pm0.03$. Seven (three) of the sources are 
individually 
detected at $>3\sigma$ ($>5\sigma$) levels. Mean millimeter fluxes are higher
for sources with the reddest mid-infrared/optical colors.
Optically faint but with relatively low mm to mid-infrared ratio, 
the typical SEDs are inconsistent with 
redshifted SED shapes of local star-forming infrared galaxies. They 
also differ from SEDs of typical 
submillimeter selected galaxies, with the 24$\mu$m sources that are 
individually detected by MAMBO possibly representing intermediate objects.
Compared to star-forming galaxies, a stronger but optically obscured 
mid-infrared component without 
associated strong far-infrared emission has to be included. This component
may be due to luminous optically obscured AGN, which would 
represent a significant part of the high redshift AGN population.

\end{abstract}


\section{Introduction}
Significant progress has been made in the last few years in detecting infrared
galaxies at high redshift and characterizing their nature, evolution, and
contribution to the cosmic background.  ISOCAM has detected at $15\,{\rm \mu
m}$ a population of $z \lesssim 1$ luminous star--forming galaxies 
\citep[e.g.][]{elbaz02}, still fairly bright in the optical 
\citep[e.g.][]{flores99}.  SCUBA and MAMBO surveys of
small fields detect a distinct $z \sim 2.5$ population of hyperluminous
star--forming and active galaxies \citep[for a review see][]{blain02} that 
are optically extremely faint \citep[e.g.][]{ivison02,dann04}.  
Spitzer Space Telescope
$24\,{\rm \mu m}$ surveys sample the redshifted, rest--frame $6-12\,{\rm \mu
m}$ strong dust emission from PAH and very small grains in galaxies at $z
\sim 1-3$.  Therefore, deep mid-IR observations using MIPS on Spitzer 
offer a
new and efficient probe of dusty galaxies at high redshift 
\citep[e.g.][]{perez05}.  Spitzer $24\mu$m surveys are expected to 
detect all of the ISOCAM and most of the
SCUBA/MAMBO sources, bridge the gap between these two populations in redshift
and luminosity, and search for entirely new categories of sources.

The first non--proprietary Spitzer survey is the First Look Survey
(FLS), covering 3.7 square degrees in seven bands.
More than 18000 sources are detected to 
the $3\sigma$ sensitivity of $110\,{\rm \mu Jy}$ at $24\,{\rm \mu m}$.
Combining deep optical, Spitzer IRAC $8\,{\rm \mu m}$, and MIPS
$24\,{\rm \mu m}$ data, \citet{yan04} have obtained the first 
characterization of a large sample of $24\,{\rm \mu m}$--selected sources.
These sources have colors
ranging from those of relatively low--redshift, fairly inactive galaxies to
extremely red $24\,{\rm \mu m}/R$ and $24\,{\rm \mu m}/8\,{\rm \mu m}$ colors
that can only be explained by luminous obscured starbursts or AGN at high
redshift. 
The available 
color information is not unique for identifying the nature of these objects, 
however, and the applicability of the local SED templates has to be critically
examined. This is particularly true for the rare and extreme 
$24\,{\rm \mu m}$--selected objects that the FLS sample includes due to its 
large size. 
In this letter we report on MAMBO 1.2mm observations of such rare 
24$\mu$m-bright but optically faint sources, providing a 
constraint on the cold dust contribution to their spectral energy 
distributions that is 
essential for characterizing their nature and their relation to other 
populations of high--redshift infrared galaxies. 

\section{Sample and Results}
\citet{yan04} used Spitzer mid-IR and ground-based optical photometry
for a first characterization of 24$\mu$m sources from the FLS. We have 
selected for 1.2mm photometry a subset of bright 
(S$_{24\mu m}>$1mJy) sources that are optically faint, as 
evidenced by a 24$\mu$m to optical `color' 
$R(24,0.7)\equiv log_{10}(\nu F_\nu (24\mu m)/\nu F_\nu (0.7\mu m))$ of at least 1 (R$>$22.5 for S$_{24\mu m}=$1mJy).
We have chosen 39 of the 145 sources meeting these criteria, and added one 
with a slightly lower R(24,0.7). We refer to these sources as `obscured 
24$\mu$m sources'. The selection covers the full color spread of 
the \citet{yan04} 
objects (Fig.~\ref{fig:yancolor}), preferring objects brightest at 24$\mu$m,
and aiming for a large overlap with FLS objects studied 
by Spitzer mid-IR spectroscopy \citep[][and in preparation]{yan05}. 
Objects were selected using the \citet{yan04} flux catalogs. Since then, 
slightly revised 24$\mu$m and 8$\mu$m fluxes (M. Lacy et al. in preparation) 
have become available which we adopt in the following discussion.

Observations were spread over the pool observing sessions at the IRAM 
30m telescope in fall/winter 2004/2005, using the 117 element version of 
the Max Planck Millimeter Bolometer (MAMBO) array \citep{kreysa98} 
operating at a wavelength of 1.2mm.
On-off observations were typically obtained in blocks of 6 scans of 20 
subscans each, and repeated in later observing nights until reaching 
noise near 0.66mJy or a 5$\sigma$ detection. The median 1.2mm noise level 
reached for the sample is 0.59mJy (range 0.35 to 0.82mJy). The data were 
reduced with standard procedures 
in the MOPSIC package developed by R.~Zylka, using the default calibration 
files for these two pool periods. Table~\ref{tab:fluxes} lists the resulting 
1.2mm fluxes and their statistical uncertainties.

\begin{figure}
\epsscale{0.6}
\plotone{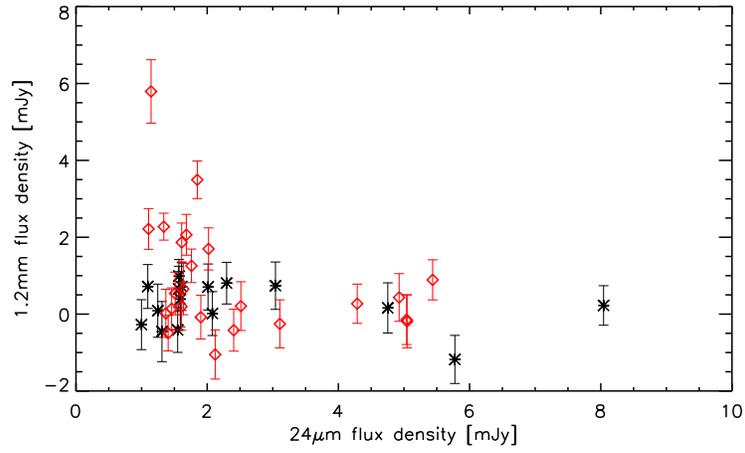}
\caption{MAMBO 1.2mm fluxes as a function of 24$\mu$m fluxes. 24$\mu$m errors 
are smaller than the symbols. Sources with R(24,0.7)$>$1.5 are shown as 
diamonds, others as asterisks.}
\label{fig:mambo24scatter}
\end{figure}

Three sources are detected individually at more than 5$\sigma$ with 1.2mm 
fluxes 
between 2.3 and 5.8mJy, and four more at 3$\sigma$ to 5$\sigma$. The total 40 
object sample is well detected statistically at a mean flux of 
0.74$\pm$0.09mJy (Table~\ref{tab:detstat}). To investigate a dependence on
the mid-IR/optical colors, we have grouped the sample in four
color `quadrants' separated at R(24,0.7)=1.5 and
 and R(24,8)=0.5. Due to the sample selection, these 
quadrants contain similar numbers of objects by design.
Table~\ref{tab:detstat} shows
clear trends in the mean 1.2mm fluxes for these subsamples. Objects redder in 
R(24,0.7) have significantly higher mean 1.2mm flux, and a trend may 
also be present with R(24,8). Values quoted in 
Table~\ref{tab:detstat} refer to means of the individual source data weighted
by the inverse of the noise squared. Very similar values are obtained 
for equal weight means. The trend in the mean fluxes agrees with the
$>3\sigma$ detection of 7/24 objects at R(24,0.7)$>$1.5 vs. 0/16 at bluer 
color. Using the formalism of \citet{stevens05}, this implies that the two
groups are different at 2.9$\sigma$ significance. The distribution of 
1.2mm vs. 24$\mu$m fluxes, 
shown in Fig.~\ref{fig:mambo24scatter}, is inconsistent with a 
single flux ratio for the entire sample, or for the subsample 
with R(24,0.7)$>$1.5. This is expected given the sample selection, and 
emphasizes that the mean 
values quoted in Table~\ref{tab:detstat} are formed over (sub)samples that 
can have noticeable spreads of intrinsic properties.
  
\section{Millimeter properties of obscured Spitzer sources}

Our color selection implies that these objects must be heavily obscured and/or
at considerable redshift. This is illustrated in Fig.~\ref{fig:yanratios},
showing that even spectral energy distributions similar to `standard' dusty 
starbursts like M82 fail to reproduce for plausible redshifts 
R(24,0.7)$>$1.2, i.e. the vast majority of our sample. R(24,0.7) will
 be yet lower for SEDs of UV-brighter starbursts like 
NGC 7714. SEDs similar to the most 
optically obscured local 
ultraluminous infrared galaxies like Arp220 reach  R(24,0.7)$>$1.5 
for redshifts around 2, but even stronger emission at 
observed 24$\mu$m (rest-frame 8$\mu$m for z$\approx$2) relative to the shorter 
wavelengths is needed to reproduce the most extreme Spitzer
objects. The same is true for the SED of the well studied less 
luminous but extremely obscured galaxy, NGC 4418. For this object, both the 
8$\mu$m and the 0.7$\mu$m flux are host dominated at z$\gtrsim$1.5. 
Varying the nucleus to host ratio moves the source 
diagonally in Fig.~\ref{fig:yanratios}. 

\begin{figure}
\epsscale{0.7}
\plotone{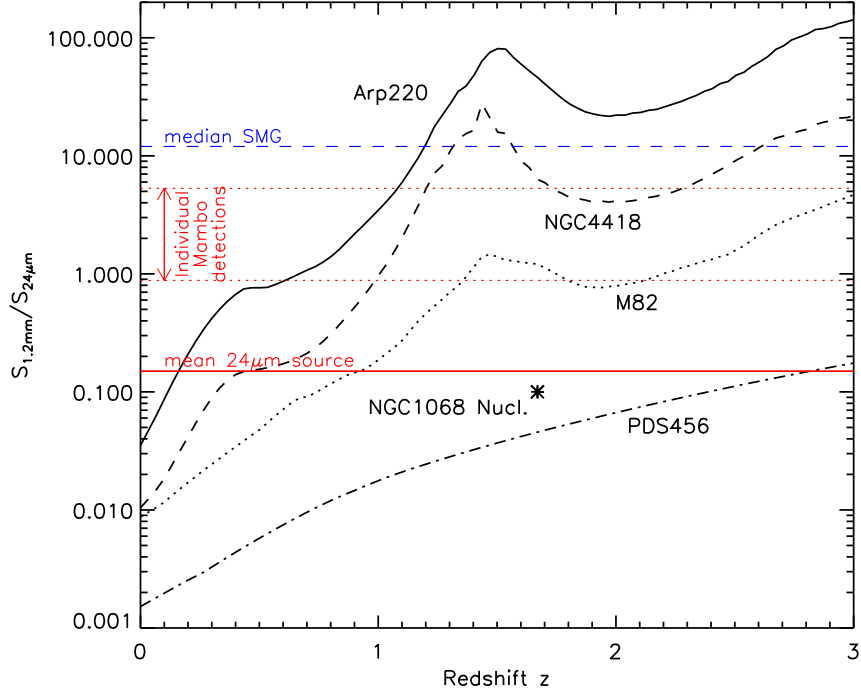}
\caption{Ratio of 1.2mm and 24$\mu$m flux density as a function of redshift
for local SED templates.  The SEDs of M82 and 
Arp220 represent typical and extremely obscured star-forming infrared 
galaxies. NGC 4418 is an extremely obscured compact object possibly hosting an 
AGN. PDS456 is a luminous quasar whose infrared SED is dominated by the AGN 
mid-IR emission, with a weak far-IR component. The AGN dominated 
nuclear region of NGC 1068 is only shown for a single redshift where the rest 
frame submm emission of the nucleus can be spatially isolated.
The horizontal continuous 
line shows the mean ratio for our sample. Horizontal dotted lines indicate
the range spanned by the targets that are individually detected by 
MAMBO. The horizontal dashed line indicates the median 1.2mm to 24$\mu$m 
flux ratio for submillimeter galaxies (SMGs) -- see text for references.
}
\label{fig:mambo24}
\end{figure}

The millimeter observations provide a crucial test of the analogy to 
local SEDs. Fig.~\ref{fig:mambo24} shows the expected ratio of 1.2mm
and 24$\mu$m flux density as a function of redshift, again for M82, Arp220, 
and NGC 4418 SEDs. The seven objects with individual MAMBO detections are 
reasonably close to the properties of these local templates. 
At $S_{1.2mm}/S_{24\mu m}\sim 1\ldots 5$, they are in a range still 
consistent with local infrared galaxy SEDs redshifted to z$\sim$2,
assuming a relatively high optical obscuration or modest additional
mid-iR emission.
In contrast, comparison of the templates to the mean values for the obscured 
Spitzer sources (Table~\ref{tab:detstat}) shows that in fact neither SED 
provides a good representation of the mean properties of our 24$\mu$m sources.
Arp 220 and NGC 4418--like SEDs, which were in a reasonable range of the 
optical/mid-IR colors, predict by far too strong 1.2mm flux for 
plausible redshifts. M82-like SEDs are closer to the observed mm to MIR ratio,
but predict too bright optical counterparts as noted above. We conclude from 
the analysis of mean properties that
the sample of obscured 24$\mu$m contains many sources that are not well 
fit by these local infrared luminous galaxy SED templates. Relative to 
the templates, they require additional rest 
frame mid-IR emission, due to a component that is both warm 
(to avoid overproducing mm emission) and optically obscured.

Another way of phrasing the need for `additional rest frame mid-IR
emission' is to state
that the overall infrared SED must be warm to hot, without a dominant 
rest frame 
far-IR peak. In a strongly simplifying single temperature blackbody 
fit, the 
observed 1.2mm to 24$\mu$m flux density ratio corresponds to a temperature
of $\sim$75K, i.e. intrinsic T$\gtrsim$200K if at z$\sim$2, and luminosities
of the order of 10$^{13}$L$_\odot$. The warm temperature is suggestive
of a strong AGN contribution. The SED of the most luminous radio-quiet 
local quasar, PDS 456 \citep{yun04} is able to fit the 1.2mm/24$\mu$m ratio 
for plausible redshifts (Fig.~\ref{fig:mambo24}), but this Type 1 object is 
too bright in the UV/optical to explain the mid-IR/optical colors for 
our sample. That more obscured AGN may be 
able to meet that constraint can be illustrated using the {\em nuclear} SED
of the classical Seyfert 2 NGC 1068. Its nuclear FIR/submm emission can 
currently be separated from the host only at 450$\mu$m with 
S$_{450\mu m}\sim 1.5$Jy \citep{papadopoulos99},  but this can be combined
with the mid-IR AGN continuum \citep{lutz00} and the rest frame UV 
taken from NED, to conclude that such an AGN SED 
would satisfy the criteria for both faintness at 1.2mm 
(Fig.~\ref{fig:mambo24}) and optical obscuration (Fig.~\ref{fig:yancolor}). 
AGN in obscured 24$\mu$m sources must suffer intermediate obscuration, 
sufficient to suppress the rest frame UV/optical but not 
moving too much energy into far-IR re-emission. Separation from host 
emission and spectral characterization of the rest frame far-IR emission
is currently incomplete for local obscured AGN, but objects like 
IRAS F00183-00183 \citep{tran01,spoon04} do have more favourable 
mid- to far-IR ratios than NGC 4418 and may fit the SEDs of some 
obscured 24$\mu$m sources.  

Low 1.2mm/24$\mu$m flux ratios can also be obtained for low redshift
(z$\lesssim $1) infrared star-forming galaxy SEDs. As already noted, this
is ruled out for the adopted templates, by the failure to meet R(24,0.7)$>$1 
at such redshifts (Fig.~\ref{fig:yanratios}). Contamination by similar sources 
may occur at the low R(24,0.7) end of the MAMBO sample if the scatter in rest 
frame SEDs extends towards smaller optical/mid-IR flux ratios. This 
scenario becomes progressively unlikely for the 
R(24,0.7)$>$1.5 sources with weak 1.2mm emission.

\citet{hunt05} use the likely compactness of star formation events in 
high redshift infrared galaxies to argue for SEDs of Blue Compact Dwarf 
Galaxies as local analogues. The most extreme case,
SBS 0335-052, shows an absorbed warm mid-IR continuum 
\citep{thuan99,houck04}. While this possible analogy has to be mentioned,
it is unclear to which extent such low metallicity dwarfs can 
relate to 
the luminous obscured 24$\mu$m sources. SBS 0335-052, in addition to its
mid-IR emission, has strong unobscured optical/UV 
emission which would place it far off the color selection of obscured
24$\mu$m sources. Rest frame submm information is not 
available to place SBS 0335-052 on Fig.~\ref{fig:mambo24}.
 
\section{Relation to other populations}

\begin{figure}
\epsscale{0.7}
\plotone{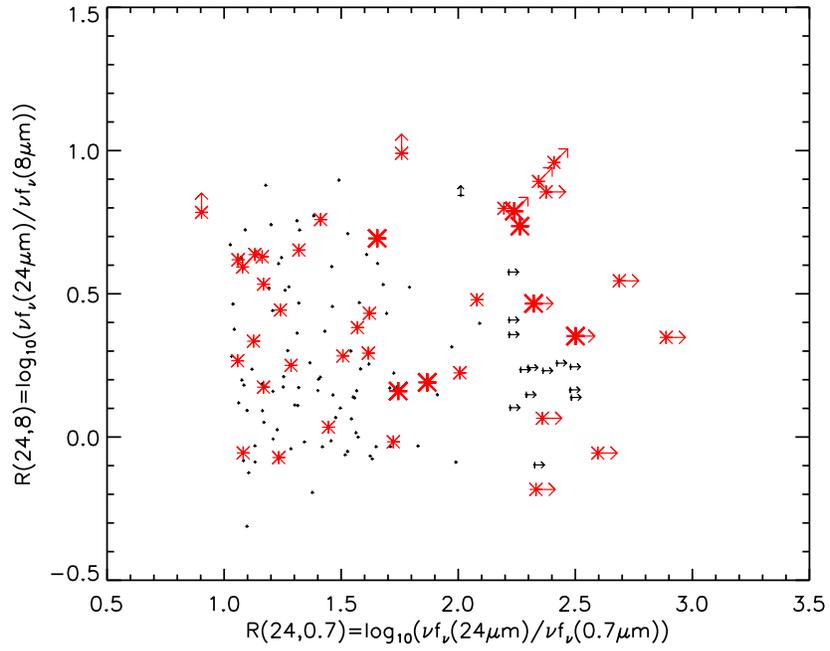}
\caption{Optical/mid-IR color-color plot of bright ($>$1mJy) 24$\mu$m 
sources from the \citet{yan04} Spitzer FLS sample, 
indicated by small crosses. Sources 
with R(24,0.7)$<$1 are not shown. Asterisks highlight the targets 
observed by MAMBO, big asterisks those individually detected by MAMBO at 
the $\geq 3\sigma$ level. 8$\mu$m and optical limits are 3$\sigma$.}
\label{fig:yancolor}
\end{figure}

\begin{figure}
\epsscale{0.7}
\plotone{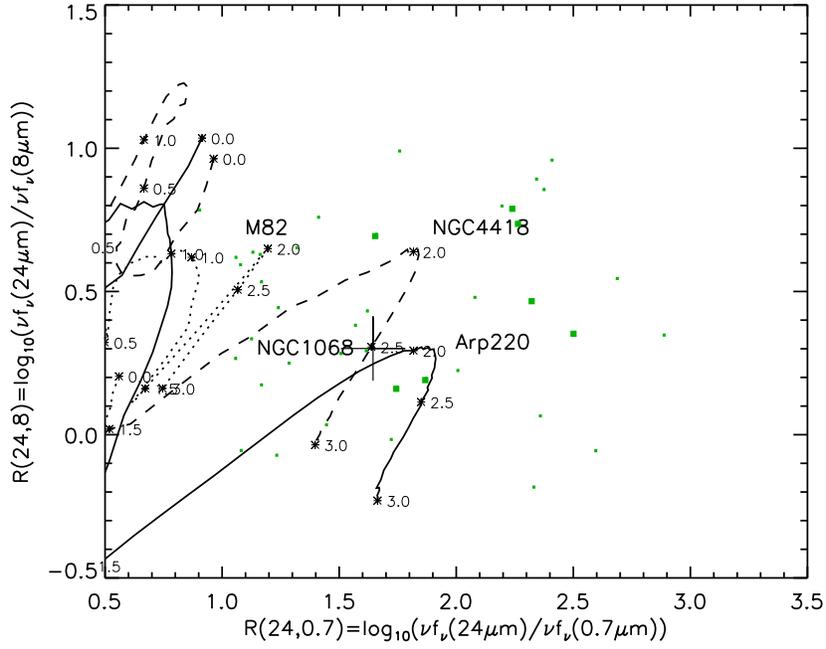}
\caption{Mid-IR/optical colors for redshifted objects with 
the SEDs of the star forming infrared galaxies M82 and Arp220, and of the 
heavily obscured galaxy NGC 4418. Locations of the FLS sources observed by 
MAMBO are repeated from Fig.~\ref{fig:yancolor}. Redshifts 
from 0 to 3 are marked in intervals of 0.5. The adopted SEDs are based on 
observations for large apertures taken from the literature. The key rest 
frame mid-IR range uses the 
ISOCAM-CVF low resolution spectra of \citet{foerster03} for M82,  O. Laurent
(priv. comm) for Arp 220, and ISOPHOT-S \citep{spoon01} and groundbased spectra
\citep{roche86} for NGC 4418. The colors of the nuclear region of the Type 2 
AGN NGC 1068 shifted to z$\sim$1.67 are indicated by a cross.}
\label{fig:yanratios}
\end{figure}

(Sub)millimeter galaxies (SMGs) are the best studied pre-Spitzer 
population of z$>1.5$ infrared galaxies, inferred to be objects at z$\sim$2.5
with roughly ULIRG like SEDs but much higher luminosity. Optical to 
mid-IR colors of bright and obscured 24$\mu$m sources 
(Figs.~\ref{fig:yancolor},\ref{fig:yanratios}) and their low surface density
compared to that of SMGs could have been consistent with 
an identification of the yet more extreme luminosity end of the SMG 
population as obscured 24$\mu$m sources. The small ratio
of 1.2mm and 24$\mu$m emission clearly argues against this being true for the 
majority of the Spitzer sources. \citet{egami04}, \citet{ivison04}, and 
\citet{frayer04} have 
reported 24$\mu$m photometry of SMGs from SCUBA and MAMBO blank field and 
radio preselected surveys. They find
much higher $S_{1.2mm}/S_{24\mu m}$, with median $\sim 12$, clearly 
inconsistent with SMGs' and obscured Spitzer sources' having similar SED 
shapes and redshifts. Again, the Spitzer objects with individual MAMBO 
detections are closer to the SMG properties.

\citet{blain04} discuss selection effects against warm SEDs in the SMG 
population and the potential of Spitzer in avoiding them. In a related 
effort, \citet{chapman04} have obtained spectroscopic redshifts of 
optically faint 
$\mu$Jy radio sources without submm detections, inferring under the
assumption of the radio -- far-IR correlation that these must represent
a population of ultraluminous star-forming z$\sim$2.2 galaxies 
characterized by 
intrinsically warm or hot dust continua. Little is known at present 
about the rest-frame mid-IR properties of these sources, making an 
assessment of the overlap with obscured 24$\mu$m sources difficult. 
\citet{frayer04} report faint 24$\mu$m fluxes or nondetections of several 
optically faint radio sources that have SCUBA 850$\mu$m flux upper 
limits of order 8mJy. If these results are representative for the full 
population
of optically faint radio sources, then their possible warm FIR SEDs cannot 
extend down into the mid-IR, and their nature must be different from the
obscured 24$\mu$m sources. If, for example, their radio emission is AGN 
boosted, their dust emission may be comparatively weak at all wavelengths.

Mid-IR spectroscopy will play a key role in identifying the 
nature of the rare obscured 24$\mu$m sources and their strong 
mid-IR emission. \citet{houck05} and \citet{yan05} have reported the
first Spitzer IRS spectroscopy of such sources, finding a predominance of 
continua and absorbed continua and indications for starburst PAH emission 
in a subset of sources. The obscured continuum spectra, which are seen only 
in a small minority of local infrared galaxies \citep{spoon02} but a 
third to half of the obscured 24$\mu$m 
sources in the subsamples already published by \citet{houck05} and 
\citet{yan05}, indicate heavily embedded energy sources that 
might be AGN. Redshift estimates, where they can be derived, suggest 
z$\sim$1.5--3. As for the photometric evidence, 
comparison to initial Spitzer spectroscopy of SMGs which shows PAHs 
\citep{lutz05} suggests that the two populations have only modest overlap,
and that there is a large contribution of warm and possibly AGN like sources
to the obscured 24$\mu$m population.

If a good part of the bright obscured 24$\mu$m sources contain luminous high z
AGN they will represent a 
significant component of the high redshift AGN population. Our MAMBO sample
is a subset of about 150 similar objects (Fig.~\ref{fig:yancolor}) 
detected over 
the 3.7 square degree FLS field. If many of these objects are 
$L_{Bol}\sim10^{13}$L$_\odot$ AGN at z$\sim$1.5-3 their comoving number 
density will be of the order $10^{-6}$Mpc$^{-3}$, comparable to that of 
similar bolometric luminosity (M$_B<-25$) optically selected quasars 
\citep{croom04}. The ongoing characterization of their redshift 
distribution and SEDs will shed more light on this issue.  

\acknowledgments
Based on observations carried out with the IRAM 30m telescope and with the 
Spitzer Space Telescope.  We thank the IRAM staff for support and advice, 
all pool observers for their help in obtaining the observations, and the 
referee for helpful comments.



\clearpage
\begin{deluxetable}{rrrrr}
\tablewidth{0pt}
\tablecaption{MAMBO 1.2mm fluxes of obscured 24$\mu$m sources
\label{tab:fluxes}}
\tablehead{
\colhead{ID}&\colhead{RA}&\colhead{DEC}&\colhead{$\rm S_{1.2mm}$}&
                                        \colhead{$\rm S_{24\mu m}$}\\
\colhead{}&\colhead{J2000}&\colhead{J2000}&\colhead{mJy}&\colhead{mJy}
}
\startdata
   39&17:17:50.65&58:47:45.1&-0.14$\pm$0.65&5.04\\
   42&17:17:58.44&59:28:16.8& 0.43$\pm$0.62&4.93\\
   45&17:11:47.46&58:58:40.0& 0.16$\pm$0.65&4.75\\
   78&17:15:38.18&59:25:40.1&-0.25$\pm$0.62&3.11\\
  110&17:12:15.44&58:52:27.9&-0.42$\pm$0.55&2.40\\
  133&17:14:33.68&59:21:19.3& 0.01$\pm$0.57&2.08\\
  180&17:15:43.54&58:35:31.2& 1.26$\pm$0.44&1.76\\
  227&17:14:56.24&58:38:16.2&-0.41$\pm$0.58&1.55\\
  289&17:13:50.00&58:56:56.8& 2.28$\pm$0.35&1.34\\
  464&17:14:39.57&58:56:32.1&-0.27$\pm$0.65&1.00\\
 7967&17:18:31.68&59:53:17.2& 0.22$\pm$0.51&8.04\\
 7977&17:14:19.96&60:27:24.8&-1.18$\pm$0.63&5.78\\
 8034&17:12:10.28&60:18:58.1& 0.74$\pm$0.62&3.04\\
 8069&17:15:00.36&59:56:11.5& 0.21$\pm$0.63&2.51\\
 8107&17:16:38.65&59:49:44.4& 0.70$\pm$0.60&2.01\\
 8135&17:14:55.71&60:08:22.6& 3.49$\pm$0.49&1.85\\
 8182&17:14:34.56&60:28:28.7& 1.87$\pm$0.50&1.61\\
 8184&17:12:26.76&59:59:53.5& 0.72$\pm$0.62&1.61\\
 8185&17:14:22.85&60:28:34.7& 0.20$\pm$0.62&1.61\\
 8196&17:15:10.28&60:09:55.2& 0.99$\pm$0.43&1.57\\
 8245&17:15:36.34&59:36:14.8&-0.49$\pm$0.46&1.41\\
 8268&17:16:24.74&59:37:52.6&-0.45$\pm$0.78&1.31\\
 8377&17:17:33.51&59:46:40.4& 0.72$\pm$0.57&1.09\\
15690&17:19:22.44&60:14:56.2& 0.89$\pm$0.52&5.44\\
15776&17:20:50.66&59:32:54.2& 0.81$\pm$0.54&2.30\\
15840&17:19:22.40&60:05:00.3&-0.08$\pm$0.57&1.90\\
15880&17:21:19.46&59:58:17.2& 2.06$\pm$0.53&1.68\\
15890&17:24:48.65&60:14:39.1& 0.65$\pm$0.67&1.64\\
15929&17:24:06.95&59:24:16.3& 0.54$\pm$0.56&1.51\\
15943&17:20:18.06&59:29:01.2& 0.14$\pm$0.55&1.46\\
15972&17:25:29.32&59:57:30.7& 0.02$\pm$0.63&1.37\\
16030&17:20:00.32&60:15:20.9& 0.09$\pm$0.69&1.25\\
22195&17:20:44.81&58:29:23.7&-0.19$\pm$0.69&5.05\\
22204&17:18:44.38&59:20:00.5& 0.27$\pm$0.51&4.29\\
22303&17:18:48.80&58:51:15.1&-1.05$\pm$0.64&2.12\\
22314&17:19:27.28&59:15:36.3& 1.70$\pm$0.55&2.02\\
22397&17:20:05.96&59:17:45.0& 0.39$\pm$0.72&1.59\\
22404&17:21:51.77&58:53:27.7& 0.66$\pm$0.59&1.57\\
22563&17:22:49.00&58:50:27.5& 5.79$\pm$0.82&1.14\\
22582&17:21:24.58&59:20:29.5& 2.22$\pm$0.53&1.11\\
\enddata
\end{deluxetable}
\clearpage

\begin{deluxetable}{cccrrr}
\tablewidth{0pt}
\tablecaption{Mean 1.2mm properties for different ranges of 
mid-IR/optical colors\label{tab:detstat}}
\tablehead{
\colhead{R(24,0.7)} & \colhead{R(24,8)} & \colhead{N} 
&\colhead{$\rm\sigma_{1.2}$} 
&\colhead{$\rm S_{1.2mm}$}&\colhead{$\rm S_{1.2mm}/S_{24\mu m}$}\\
\colhead{} &\colhead{} &\colhead{} &\colhead{mJy} &\colhead{mJy} &\colhead{}
}
\startdata
all      & all     & 40 &0.59& 0.74$\pm$0.09&0.15$\pm$0.03\\
$<$1.5   &$<$0.5   &  8 &0.65& 0.15$\pm$0.22&0.01$\pm$0.05\\
$<$1.5   &$\geq$0.5&  8 &0.59& 0.46$\pm$0.20&0.26$\pm$0.11\\
$\geq$1.5&$<$0.5   & 15 &0.57& 0.91$\pm$0.14&0.18$\pm$0.05\\
$\geq$1.5&$\geq$0.5&  9 &0.53& 1.08$\pm$0.18&0.52$\pm$0.10\\
\enddata
\tablecomments{See Fig.~\ref{fig:yancolor} for the distribution of 
sources within the R(24,0.7) and R(24,8) colors. $\rm\sigma_{1.2}$ 
indicates the median 1.2mm flux error for the sources in each bin.}
\end{deluxetable}

\clearpage

\clearpage

\clearpage

\clearpage

\clearpage


\begin{thebibliography}{}
\bibitem[Blain et al.(2002)]{blain02} Blain, A.W., Smail, I., Ivison, R.J., 
  Kneib, J.-P., Frayer, D.T., 2002, PhR, 369, 111
\bibitem[Blain et al.(2004)]{blain04} Blain, A.W., Chapman, S.C., Smail, I.,
Ivison, R. 2004, \apj, 611, 52
\bibitem[Chapman et al.(2004)]{chapman04}Chapman, S.C., Smail, I., 
Blain, A.W., Ivison, R.J. 2004, \apj, 614, 671 
\bibitem[Croom et al.(2004)]{croom04} Croom S.M., Smith, R.J., Boyle, B.J.,
Shanks, T., Miller, L., Outram, P.J., Loaring, N.S. 2004, \mnras, 249, 1397  
\bibitem[Dannerbauer et al.(2004)]{dann04} Dannerbauer, H., Lehnert, M.D., 
Lutz, D., Tacconi, L., Bertoldi, F., Carilli, C., Genzel, R., Menten, K. 2004,
\apj, 606, 664
\bibitem[Egami et al.(2004)]{egami04} Egami, E., et al. 2004, \apjs, 154, 130
\bibitem[Elbaz et al.(2002)]{elbaz02} Elbaz, D., Cesarsky, C.J., Chanial, P.,
Aussel, H., Franceschini, A., Fadda, D., Chary, R.R. 2002, \aap, 384, 848
\bibitem[Flores et al.(1999)]{flores99} Flores, H., et al. 1999, \apj, 517, 148
\bibitem[F\"orster Schreiber et al.(2003)]{foerster03} F\"orster Schreiber, 
N.M., Sauvage, M., Charmandaris, V., Laurent, O., Gallais, P., Mirabel, I.F., 
Vigroux, L. 2003, \aap, 399, 833
\bibitem[Frayer et al.(2004)]{frayer04} Frayer, D.T., et al. 2004, \apjs, 
154, 137
\bibitem[Houck et al.(2004)]{houck04} Houck, J., et al. 2004, \apjs, 154, 211 
\bibitem[Houck et al.(2005)]{houck05} Houck, J., et al. 2005, \apj, 622, L105 
\bibitem[Hunt et al.(2005)]{hunt05} Hunt, L., Maiolino, R. 2005, \apj, 
626, L15
\bibitem[Ivison et al.(2002)]{ivison02} Ivison, R.J., et al. 2002, \mnras,
337, 1
\bibitem[Ivison et al.(2004)]{ivison04} Ivison, R.J., et al. 2004, 
\apjs, 154, 124
\bibitem[Kreysa et al.(1998)]{kreysa98} Kreysa, E. et al.\ 1998, \procspie,
3357, 319
\bibitem[Lutz et al.(2000)]{lutz00} Lutz, D., Sturm, E., Genzel, R., 
Moorwood, A.F.M., Alexander, T., Netzer, H., Sternberg, A. 2000, \apj, 536, 
697   
\bibitem[Lutz et al.(2005)]{lutz05} Lutz, D., Valiante, E., Sturm, E., 
Genzel, R., Tacconi, L.J., Lehnert, M.D., Sternberg, A., Baker, A.J. 2005,
\apj, 625, L83
\bibitem[Papadopoulos \& Seaquist(1999)]{papadopoulos99} Papadopoulos, P.P.,
Seaquist, E.R. 1999, \apj, 514, L95
\bibitem[P\'erez Gonz\'alez et al.(2005)]{perez05} P\'erez Gonz\'alez, P.G., 
et al. 2005, \apj, in press (astro-ph/0505101)  
\bibitem[Roche et al.(1986)]{roche86} Roche, P.F., Aitken, D.K., Smith, C.H.,
James, S.D. 1986, \mnras, 218, 19p
\bibitem[Spoon et al.(2001)]{spoon01} Spoon, H.W.W., Keane, J.V., 
Tielens, A.G.G.M., Lutz, D., Moorwood, A.F.M. 2001, \aap, 365, L353
\bibitem[Spoon et al.(2002)]{spoon02} Spoon, H.W.W., et al. 2002, \aap,
385, 1022
\bibitem[Spoon et al.(2004)]{spoon04} Spoon, H.W.W., et al. 2004,
\apjs, 154, 184
\bibitem[Stevens et al.(2005)]{stevens05} Stevens, J.A., Page, M.J.,
Ivison, R.J., Carrera, F.J., Mittaz, J.P.D., Smail, I., McHardy, I.M. 2005
\mnras, 360, 610
\bibitem[Thuan et al.(1999)]{thuan99} Thuan, T.X., Sauvage, M., Madden, S.
1999, \apj, 516, 783
\bibitem[Tran et al.(2001)]{tran01} Tran, Q.D., et al. 2001, \apj, 552, 527
\bibitem[Yan et al.(2004)]{yan04} Yan, L., et al. 2004, \apjs, 154, 60
\bibitem[Yan et al.(2005)]{yan05} Yan, L., et al. 2005, \apj, 628, 604
\bibitem[Yun et al.(2004)]{yun04} Yun, M.S., Reddy, N.A., Scoville, N.Z., 
Frayer, D.T., Robson, E.I., Tilanus, R.P.J. 2004, \apj, 601, 723  
\end{thebibliography}
\end{document}